\documentstyle[11pt,rafringe,twoside,graphicx,makeidx]{article}
\makeindex
\makeindex

\def\edcomment#1{\iffalse\marginpar{\raggedright\sl#1\/}\else\relax\fi} \marginparwidth 1.25in
\reversemarginpar

\begin{document} 
\title{Self-Stark Limit on Brightness Temperature in Cosmic Masers} 
\author{V. I. Slysh} 
\affil{Arecibo Observatory, HC3 Box 53995, Arecibo PR 00612}
\affil{Astro Space Center, Profsoyuznaya st. 84/32,
117997 Moscow, Russia}

\begin{abstract} 
The radiation broadening of narrow maser lines by its own emission which can also be described as resonance
Self-Stark effect
puts an upper limit on the brightness temperature of cosmic masers. It is shown that available 
measurements of the brightness temperature of OH and water masers are consistent with the observed
narrow line widths only if the maser emission is highly directive.
\end{abstract}

\section{Introduction} 

Methanol, OH and water masers are generally seen toward star-forming regions and exhibit narrow, strong emission
line features coming from bright small spots. The brightness temperature $T_{\rm b}$ of the spots is very high, from $10^9$\,K
for methanol masers, to $6\times10^{12}$\,K for OH masers, and $10^{15}$\,K for water masers. In many cases the above values
are only lower limits since maser spots are barely resolved with VLBI. Variability time scale often indicates, in an
indirect way, a smaller linear size and still higher brightness temperature. There must be physical limits on the 
brightness temperature as there are for  thermal and synchrotron emission in continuum sources. The brightness 
temperature of the thermal free-free emission from ionized gas depends on the kinetic temperature $T_{\rm e}$ of electrons
and optical depth. It is the highest for optically thick sources and is equal to the kinetic temperature. In 
H\,{\sc ii} regions the kinetic temperature of electrons is limited to $T_{\rm e}$=10,000\,K by the energy losses from 
electrons through the line emission of heavy elements. Similarly, in synchrotron sources the brightness temperature is
at maximum in optically thick sources, and is equal to the kinetic temperature of relativistic electrons.
The kinetic energy of relativistic electrons is limited to $10^{11}$--$10^{12}$\,K by the energy loss through
inverse Compton emission on synchrotron photons (Kellermann \& Pauliny-Toth 1969). 
The same arguments can be applied to molecular line emission: in the optically thick sources the brightness
temperature in the line is limited by the gas kinetic temperature. The molecular maser sources make an exception
because the absorption coefficient and optical depth are both negative due to the population inversion: the
brightness temperature of the maser line emission may increase indefinitely with the optical depth. However other mechanisms
can limit the brightness temperature of the maser lines, such as the radiation line broadening, or the resonance Stark-effect
caused by the maser radiation field (we call it Self-Stark limit by analogy with Self-Compton limit).

\section{Radiation, or resonance Stark broadening of maser lines}

The natural band width of a molecular emission line in the absence of external field is $\Delta\nu=A/2\pi$,
where $A=64\pi^4\nu_0^3\mu^2/3hc^3$ is spontaneous emission rate at the frequency of the line $\nu_0$, $\mu$ is the
dipole-moment matrix element for the transition, and $h$ is Planck constant. This is a very small band width
compared to the observed line widths. For OH line at 1665\,MHz, $A=7.1\times10^{-11}$\,s$^{-1}$, and for water line at 22.235\,GHz, $A=2\times10^{-9}$\,s$^{-1}$, and the natural band widths are $10^{-11}$\,Hz and $3\times10^{-10}$\,Hz,
respectively. In strong masers
like \Index{W3\,(OH)} or \Index{W\,49} every OH or H$_2$O molecule is exposed to the intense radiation field, produced by all other molecules in the maser
emission region. That causes frequent up and down transitions between the energy levels involved, due to induced emission and absorption.
The time which a molecules spends on the energy level is greatly reduced, by a factor $2kT_{\rm b}/h\nu_0$. The linewidth is increased by
the same factor which can be very large when the brightness temperature is high. For OH maser with brightness temperature $6\times10^{12}$\,K
the linewidth is 1,700\,Hz (0.3\,km\,s$^{-1}$), and for water maser with the brightness temperature $10^{15}$\,K it is 0.6\,MHz (8\,km\,s$^{-1}$).
This linewidth is comparable to or even greater than observed line widths in masers.
Einstein induced emission and absorption coefficients can be used only in the weak field case. For the strong field which is the case in water
masers a correct description of the interaction between a molecule and the narrow band radiation field is the resonance Stark effect approach.
In the weak field the resonance Stark effect is quadratic with respect to the electric field strength $E$, and causes the line broadening. In the strong field the resonance Stark effect becomes linear, and causes line splitting into a main line at the transition frequency $\nu_0$ and two
satellite lines at frequencies $\nu_0$$\pm$$\mu$$E/h$  for non-degenerate energy levels such as are levels of the OH line transition at 1665\,MHz. The energy levels of the H$_2$O transition at 22.235\,GHz are degenerate, and are split in several satellite components (Slysh 1973). AC electric field  strength $E$ is related to the brightness temperature $T_{\rm b}$ of the maser radiation of band width $\delta\nu$ as $E=8\pi(kT_{\rm b})^{0.5}\nu_0(\delta\nu)^{0.5}c^{-1.5}$. This relation holds
for the isotropic distribution of the radiation intensity. If the radiation is concentrated in the solid angle $\Omega$, the electric field strength must be multiplied by $(\Omega/4\pi)^{0.5}$, with the corresponding reduction of the line splitting. For the line broadening
the reduction factor is $\Omega/4\pi$.

\begin{figure}[htb!]
\begin{center}
\includegraphics[clip,bb=40 105 570 670,width=0.80\textwidth]{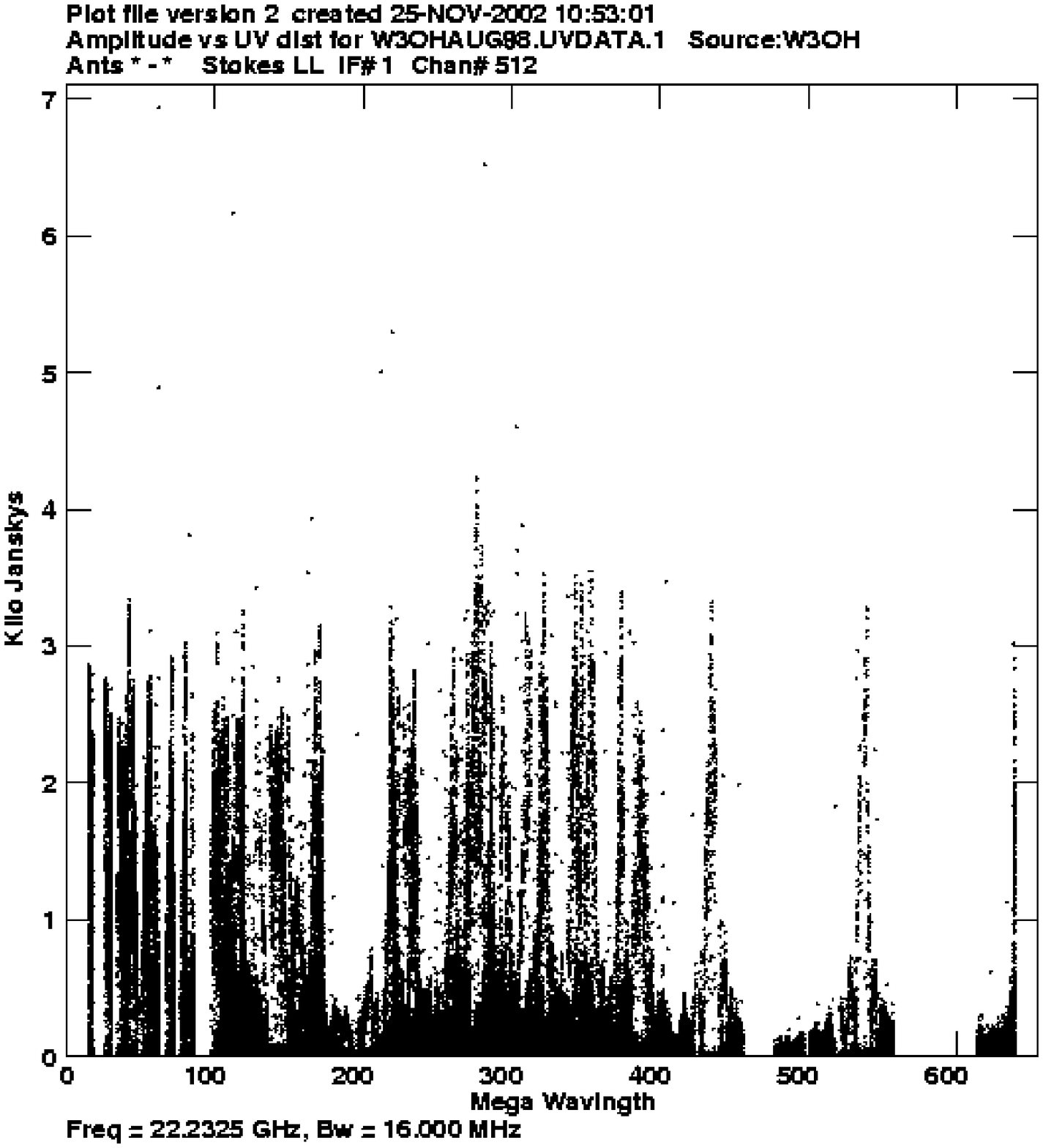}
\end{center}
\vspace*{-15pt}
\caption{Correlated flux vs baseline for the strongest H$_2$O maser spectral feature in W3OH obtained from VLBA archive data}
\end{figure}

\begin{figure}[htb!]
\begin{center}
\includegraphics[clip,bb=40 105 570 670,width=0.85\textwidth]{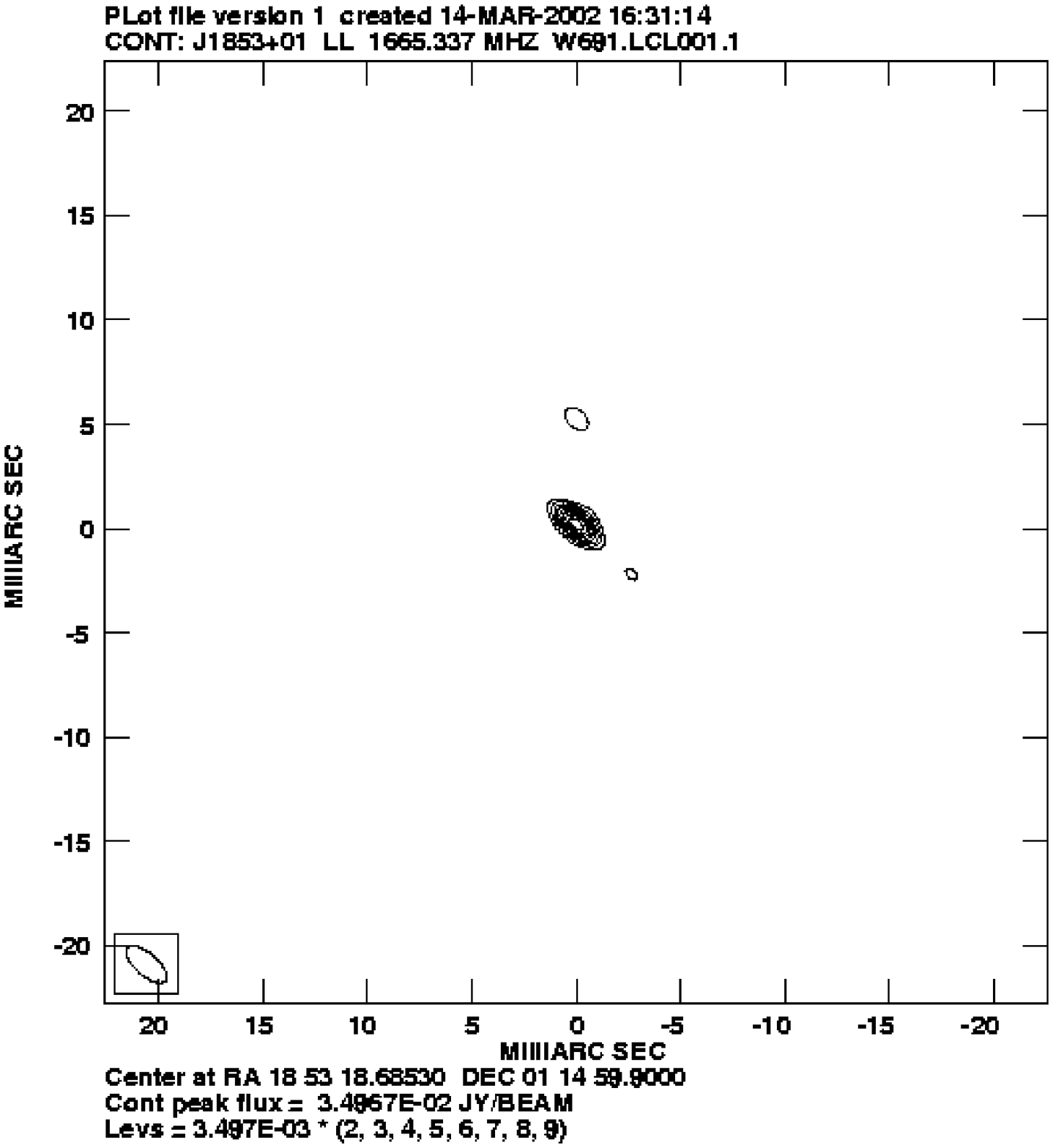}
\end{center}
\vspace*{-15pt}
\caption{A map of the most compact OH-maser emission spectral feature in \Index{OH\,34.26+0.15}. Participating telescopes were phased VLA, 70\,m 
Tidbinbilla and 8\,m space telescope \textsl{HALCA}. Synthesized beam was 2.4$\times$1.1\,mas}
\end{figure}

\section{Brightness temperature of H$_2$O and OH masers}

The strongest spectral feature in H$_2$O maser \Index{W3\,OH} is completely unresolved on all VLBA baselines (Fig.\ 1), and its brightness temperature
exceeds $4\times10^{15}$\,K. Another strong H$_2$O maser, an outburst in 
\Index{Orion-KL}, was measured  with the space radio telescope \textsl{HALCA} to have brightness temperature
$T_{\rm b}$=$10^{16}$\,K (Kobayashi et al 2000). OH maser \Index{OH\,34.26+0.15} was unresolved on baselines from ground to space radio telescope \textsl{HALCA}, and 
the brightness temperature exceeds $4\times10^{12}$\,K (Fig.~2).

\section{Directivity of the maser emission}

No line splitting due to the resonance Stark effect was observed in H$_2$O or OH masers. Also, no significant radiation line broadening
is seen in the line profiles. This means that the electric field strength in the masers is weak enough in spite of the very high 
brightness temperature, because of the small solid angle in which emission is directed. Assuming that the the radiation broadening
contributes to the total linewidth less than 1/5 of the linewidth, one can put an upper limit on the solid angle.
One has for OH maser \Index{OH\,34.26+0.15} an upper limit for the emission solid angle $\Omega/4\pi$=0.02.
For H$_2$O masers in \Index{W3\,OH} one has $\Omega/4\pi$=0.005, and for the outburst H$_2$O maser in \Index{Orion-KL} $\Omega/4\pi$=0.002 as the
upper limits.

\section{Conclusion}

The brightness temperature of cosmic masers is limited by the radiation broadening in the intense radiation field at the frequency
of the maser emission. Up to now no radiation broadening has been observed in OH and H$_2$O masers, which can be due to the
very small solid angle in which the emission is directed. The theory of the resonance radiation broadening allows us to put stringent
upper limits on the emission solid angle. This may be important for the further development of cosmic maser models.

\printindex
\end{document}